# Symmetries In Evolving Space-time and Their Connection To High-frequency Gravity Wave Production

A.W. Beckwith[1]

[1]APS
abeckwith@UH.edu

**Abstract.** We claim that linking a shrinking prior universe via a wormhole solution for a pseudo-time-dependent Wheeler-De Witt equation permits the formation of a short-term quintessence scalar field that is tied to an initial configuration of the Einstein field equations. Symmetries allow for creating high-frequency gravity waves at the onset of inflation. Our construction is a way of keeping a large, but finite number of computations for the evolution of the universe intact during cosmological expansion of the volume of our present universe, assuming Seth Lloyd's model of the universe as a quantum computer has validity. The wormhole acts as a thermal bridge between prior and present universes and is less than Planck's time duration, yet has consequences leading to our present cosmological era.

## INTRODUCTION

We begin first with a restatement of the physics leading to a wormhole solution for transfer of vacuum energy from a prior universe to today's expanding universe. We then argue that such a vacuum-energy transfer is a necessary condition for forming short-term quintessence scalar fields. This allows us to form quintessence scalar-field behavior that is consistent with the $w = -1$ value for the ratio of pressure and density, which is well within the known red shift $z = 1100$ cosmic microwave background barrier. We should note that initially, $w = 0$ at the initial onset of the big bang. Also that the CMB radiation cutoff region is 380 to 400 thousand years after the big bang. At that time, we are able to start getting a separation of photons and matter, so we can observe stellar astrophysical processes reliably. After restating how a wormhole solution forms, the results heavily depend on the scalar field turning into a real quality, after the initial phases of inflation, as well as rapidly damping out as the vacuum energy creates emergent field conditions, allowing for the relic production of gravitons. The existence of the solution of the Wheeler-DeWitt equation with a pseudo-time-like component provides an additional symmetry to space-time evolution, which is broken by the chaotic regime of the scale factor.

This leads to a bifurcation in the evolution of the quintessence scalar field that comes from initially zero value for the scalar field in the initial onset of inflation. In the present day era, the cosmological data set leads us to conclude that the rate of expansion of the universe is actually increasing. In between these two zero values of the scalar field, we have non-zero values of the scalar field. This, with the Seth Lloyd model of the universe as a quantum computing device, permits us to specify a shift to high-frequency gravity waves. This is a way to keep a finite but large number of computational bits of information for modeling universe expansion, regardless of how large the universe becomes in the present to far future.

## HOW A WORMHOLE FORMS

Lorentzian wormholes have been modeled quite thoroughly. Visser (1995) states that the wormhole solution does not have an event horizon hiding a singularity, i.e., there is no singularity in the wormhole that is held open by dark energy. We are presenting a wormhole as a bridge between a prior to a present universe, as Crowell (2005) refers to in his reference on quantum fluctuations of space-time. The equation for thermal/vacuum energy flux, which leads to a wormhole, uses a pseudo-time-like space coordinate in a modified Wheeler-De Witt equation for a bridge between two universes. We also state that the wormhole solution is dominated by a huge vacuum-energy value.

To show this, we use results from Crowell (2005) on quantum fluctuations in space-time. This gives a model from a pseudo-time-component version of the Wheeler-De Witt equation, with use of the Reinssner-Nordstrom metric to help us obtain a solution that passes through a thin shell separating two space-times. The radius of the shell, $r_0(t)$ separating the two space-times is of length $l_P$ in approximate magnitude, leading to a domination of the time component for the Reissner – Nordstrom metric

$$dS^2 = -F(r) \cdot dt^2 + \frac{dr^2}{F(r)} + d\Omega^2 \qquad (1)$$

This has:

$$F(r) = 1 - \frac{2M}{r} + \frac{Q^2}{r^2} - \frac{\Lambda}{3} \cdot r^2 \xrightarrow{T \to 10^{32} Kelvin \sim \infty} -\frac{\Lambda}{3} \cdot (r = l_P)^2 \qquad (2)$$

This assumes that the cosmological vacuum energy parameter has a temperature-dependence as outlined by Park (2003), leading to

$$\frac{\partial F}{\partial r} \sim -2 \cdot \frac{\Lambda}{3} \cdot (r \approx l_P) \equiv \eta(T) \cdot (r \approx l_P) \qquad (3)$$

a wave-functional solution to a Wheeler-De Witt equation bridging two space-times. This solution bridging two space-times is similar to that being made between these two space-times with "instantaneous" transfer of thermal heat, as given by Crowell (2005)

$$\Psi(T) \propto -A \cdot \{\eta^2 \cdot C_1\} + A \cdot \eta \cdot \omega^2 \cdot C_2 \qquad (4)$$

This has $C_1 = C_1(\omega,t,r)$ as a cyclic and evolving function in terms of frequency, time, and spatial function, with the same thing describable about $C_2 = C_2(\omega,t,r)$, with $C_1 = C_1(\omega,t,r) \neq C_2(\omega,t,r)$. The upshot of this is that a thermal bridge exists between a shrinking prior universe collapsing to a singularity and an expanding universe expanding from a singularity, with an almost instantaneous transfer of heat from that prior universe to today's cosmology. The thermal bridge being modeled as a wormhole is a necessary and sufficient condition for an almost instantaneous transferal mechanism of matter-energy to today's universe from a prior cosmological collapse. We get this by assuming that the absolute value of the five-dimensional "vacuum state" parameter varies with temperature T, as Beckwith (2007) writes

$$|\Lambda_{5-dim}| \approx c_1 \cdot (1/T^\alpha) \qquad (5)$$

in contrast with the more traditional four-dimensional version, minus the minus sign of the brane-world theory version. The five-dimensional version is based on brane theory and higher dimensions, whereas the four-dimensional version is linked to more traditional de Sitter space-time geometry, as given by Park (2003)

$$\Lambda_{4-dim} \approx c_2 \cdot T^\beta \qquad (6)$$

If we look at the range of allowed upper bounds for the cosmological constant, there exists a difference in values between what Park (2003) predicts--an almost infinite value--to a much lower value given by Barvinsky (2006), which is 360 times the square of Planck's mass. The difference in these two values is commensurate with the existence of a symmetry-breaking phase transition, where we predict spin two gravitons are released, and also when we observe axion domain wall decay. Specifies as an upper limit is based on thermal input that a phase transition is occurring at or before Planck's time. This allows for a brief interlude of quintessence. We should note that this assumes that a release of gravitons occurs, which leads to a removal of graviton-energy-stored contributions to this cosmological parameter, with $m_P$ as the Planck mass, i.e., the mass of a black hole of "radius" on the order of magnitude of Planck length $l_P \sim 10^{-35}$ m.

This leads to Planck's mass $m_P \approx 2.17645 \times 10^{-8}$ kilograms, as alluded to by Barvinsky (2006):

$$\Lambda_{4-\dim} \propto c_2 \cdot T \xrightarrow{graviton-production} 360 \cdot m_P^2 << c_2 \cdot [T \approx 10^{32} K]. \tag{7}$$

Right after the gravitons are released one still sees a drop-off of temperature contributions to the cosmological constant. Then we can write, for small time values, $t \approx \delta^1 \cdot t_P, 0 < \delta^1 \leq 1$, and for temperatures sharply lower than $T \approx 10^{12} Kelvin$, as commented on by Beckwith (2007), that there exists a positive integer $n$, which leads to a sharp phase-transition drop for temperature values as we approach Planck's time interval, $t_P \sim 10^{-44}$ seconds.

After the exit of vacuum "thermal" energy from the mouth of the wormhole bridge between a prior universe and our universe, within a Planck time interval, we observe a *decrease* in values of the cosmological constant in the four-dimensional world and an *increase* in the absolute value of the "vacuum energy" in the five-dimensional brane world. However, their absolute magnitudes are approximately the *same* after the Planck time interval.

At and before this region of relative equivalence of these magnitudes (the cosmological constant in four space and the absolute value of vacuum energy in five dimensional brane worlds), a scale-factor discontinuity region (referred to later in this paper) exists. In the region of time and space before we traverse this thermal/spatial scale factor discontinuity region, we have $10^{10}$ or so bits of "information."

After traversing this thermal/spatial scale-factor discontinuity region, it rapidly increase to $10^{120}$ or so bits of information, due to an increase in complexity of the cosmological space-time structure. The dividing line between these two regions of complexity is shown by the behavior of the four-dimensional cosmological constant energy and the five dimensional "vacuum energy" expressions discussed above.

Note that brane world physics completely breaks down before the thermal/spatial scale-factor discontinuity barrier during the Planck time interval. If we take this picture literally, we run up against nonphysical situations, such as having both tiny five-dimensional "brane world" vacuum energy and huge four-dimensional cosmological constant energy in the neighborhood of the mouth of the worm hole. What we do state, however, is that in the region *before* this discontinuity region, we have less than or equal to $10^{10}$ bits of "information." This changes to a four-dimensional "cosmological constant" energy far smaller than the magnitude of the five-dimensional brane word vacuum energy, becoming $10^{120}$ bits of actual cosmological information and leading to the large-scale creation of structure in our universe.

The time interval at this crossing of values of the brane world absolute magnitude of the cosmological vacuum energy parameter with that of the four dimensional cosmological parameter is roughly within an order of magnitude of the same time of one Planck time $t_P \sim 10^{-44}$ seconds interval of time "distance" from the initial nucleation of influx of initial space-time matter-energy from the wormhole source. We assume that during this process, we have a drop in temperatures from $T \approx 10^{32}$ Kelvin, which is what we would get as a matter-energy vacuum-energy state emerges from the mouth of the Lorentzian wormhole to

$T \approx 10^{12}$ Kelvin. This is a huge drop in temperature at the same time that relic gravitons are released. The initial temperature is in the range of needed thermal excitation levels required for quantum-gravity processes to be initiated at the onset of a new universe nucleation.

$$\frac{\Lambda_{4-dim}}{|\Lambda_{5-dim}|} - 1 \approx \frac{1}{n} \qquad (8)$$

The transition outlined in Eqn. (7) above has a starting point with extremely high temperatures created by a vacuum-energy transfer between a prior universe and our present universe, as outlined by Eqn. (3) and Eqn. (4) above. Whereas the regime where we look at an upper bound to vacuum energy in four dimensions is outlined in Eqn. (8) above. This wormhole solution is a necessary and sufficient condition for thermal transfer of heat from that prior universe to allow for graviton production under relic inflationary conditions.

## Claim 1: The Following Are Equivalent

1. There exists a Reisnner-Nordstrom Metric with -F(r) dt² dominated by a cosmological vacuum energy term, $(-\Lambda/3)$ times $dt^2$, for early-universe conditions in the time-range less than or equal to Planck's time $t_P$.
2. A solution for a pseudo-time-dependent version of the Wheeler-De Witt equation exists with a wave function $\Psi(r,t,T)$, forming a wormhole bridge between two universe domains, with $\Psi(r,t,T) = \Psi(r,-t,T)$ for a region of space-time before signal causality discontinuity, and for times $|t| < t_P$.
3. The heat flux-dominated vacuum energy value given by $\Psi(r,t,T)$ contributes to a relic graviton burst, in a region of time less than or equal to Planck's time $t_P$.

The proof of claim 1 is referenced via an article, Beckwith (2007). This claim establishes the structure we outline as to our causal discontinuity approach to wormholes.

## PRESENTING EVIDENCE FOR CAUSAL DISCONTINUITY DUE TO THE TRANSFER OF THERMAL-BASED VACUUM ENERGY IMPLIED BY THE WHEELER-DE WITT EQUATION WORMHOLE SOLUTION

Begin first by presenting a version of the Friedmann equation given by Peter Frampton:

$$(\dot{a}/a)^2 = \frac{8\pi G}{3} \cdot [\rho_{rel} + \rho_{matter}] + \frac{\Lambda}{3} \qquad (9)$$

We argue that the existence of such a nonlinear equation for early-universe scale-factor evolution introduces a de facto "information" barrier between a prior universe, which, as we argue, can only include thermal bounce input to the new nucleation phase of our present universe. To see this, we can turn to Dr.

Dowker's paper on causal sets [1]. These require the following ordering with a relation $\prec$, where we assume that initial relic space time is replaced by an assembly of discrete elements, so as to create, initially, a partially ordered set $C$:

(1) If $x \prec y$, and $y \prec z$, then $x \prec z$.

(2) If $x \prec y$, and $y \prec x$, then $x = y$ for $x, y \ \varepsilon \ C$.

(3) For any pair of fixed elements $x$ and $z$ of elements in $C$, the set $\{y \mid x \prec y \prec z\}$ of elements lying in between x and z is finite.

Items (1) and (2) give us that we have $C$ as a partially ordered set and the third item permits local finiteness. When combined with as a model for how the universe evolves via a scale factor equation, this permits us to write, after we substitute $a(t^*) < l_P$ for $t^* < t_P =$ Planck time, and $a_0 \equiv l_P$, and $a_0/a(t^*) \equiv 10^\alpha$ for $\alpha >> 0$ into a discrete equation model of Eqn (5) leads to:

## Claim 2: Using the Friedmann Equation For the Evolution Of a Scale Factor $a(t)$, We Have a Non-Partially Ordered-Set Evolution Of the Scale Factor With Evolving Time, Implying a Causal Discontinuity.

We establish the validity of this formalism by rewriting the Friedman equation as follows**:**

$$\left[\frac{a(t^* + \delta t)}{a(t^*)}\right] - 1 < \frac{(\delta t \cdot l_P)}{\sqrt{\Lambda/3}} \cdot \left[1 + \frac{8\pi}{\Lambda} \cdot \left[(\rho_{rel})_0 \cdot 10^{4\alpha} + (\rho_m)_0 \cdot 10^{3\alpha}\right]\right]^{1/2} \xrightarrow[\Lambda \to \infty]{} 0 \quad (10)$$

So in the initial phases of the big bang, with a very large vacuum energy, we obtain the following relation, which violates (signal) causality. This is for any given fluctuation of time in the "positive" direction:

$$\left[\frac{a(t^* + \delta t)}{a(t^*)}\right] < 1 \qquad (11)$$

We argue that the existence of such a violation of causality, as shown in the evolution of the scale factor (as given in the ratio of a scale factor at a later time divided by the same factor at a prior time). The fact that the scale factor is less than one argues for a break in information propagation from a prior universe to our present universe. We have just proved non-partially ordered set evolution by deriving a contradiction from the partially ordered set assumption.

One relevant area of inquiry to be investigated in the future is the following: Is this argument valid if there is some third choice of set structure? (For instance, do self-referential sets fall into one category or another?) The answer to this, we think, lies in (entangled?) vortex structure of space-time similar to that generated in the laboratory by Ruutu (1996). Self-referential sets may be part of the generated vortex structure and we will endeavor to find if this can be experimentally investigated. If the causal-set argument, and its violation via this procedure, holds, we have the view that what we are seeing is a space-time "drum" effect, with the causal discontinuity forming the head of a "drum" for a region of about $10^{10}$ bits of

"information" before our present universe up to the instant of the big bang itself for a time region less than $t \sim 10^{-44}$ seconds in duration, with a region of increasing bits of "information" going up to $10^{120}$ due to vortex filament condensed matter style forming through a symmetry breaking phase transition. We think that elaboration and extension of the Ruutu (1996) experiments may be able to show this sort of formation of structure. They may also be part of a traversable wormhole effect between two universes, per Barcelo, and Visser (1999), once we get falsifiable criteria in place to prove the drum-head-effect model cited above. We also think our goal should be to eventually try to confirm the Visser (2002) updated version of the Sakarov (1967) model of emergent gravity. We view this as a possibility, if some of the wormhole structure that we are conjecturing between two universes is confirmed with falsifiable experimental criteria. This neatly complements the transfer of an extremely large temperature-based vacuum energy via a wormhole. This wormhole transfer from a prior universe to relic conditions in today's universe is itself an argument for causal signal breakage, since it would be effectively a faster-than-light propagation from a prior collapsing universe to today's universe.

## SETH LLOYD'S UNIVERSE AS A QUANTUM COMPUTER MODEL, WITH MODIFICATIONS

We use the formula given by Seth Lloyd (2002) that defines the number of operations the "Universe" can "compute" during its evolution. Lloyd (2002) uses the idea attributed to Landauer that the universe is a physical system with information processed over its evolutionary history. Lloyd also cites a prior paper where he attributes an upper bound to the permitted speed a physical system can have in performing operations in lieu of the Margolis/ Levitin theorem. He specifies a quantum mechanically given upper limit value (assuming E is the average energy of the system above a ground state value), obtaining a **first limit** of a quantum mechanical average energy bound value of

$$[\#operations/\sec] \leq 2E/\pi\hbar \qquad (12)$$

The **second limit** to this number of operations is strictly linked to entropy, due to considerations of limits to memory space, which Lloyd writes as

$$[\#operations] \leq S(entropy)/(k_B \cdot \ln 2) \qquad (13)$$

The **third limit,** based on strict considerations of a matter-dominated universe, relates the number of allowed computations (operations) within a volume for the alleged space of a universe (horizon). Lloyd identifies this space-time volume as $c^3 \cdot t^3$, with $c$ the speed of light, and $t$ an alleged time (age) for the universe. We further identify $E(energy) \sim \rho \cdot c^2$, with $\rho$ as the density of matter, and $\rho \cdot c^2$ as the energy density (unit volume). This leads to

$$[\#operations/\sec] \leq \rho \cdot c^2 \times c^3 \cdot t^3 \qquad (14)$$

We then can write this, if $\rho \sim 10^{-27} \, kil/meter^3$ and time as approximately $t \sim 10^{10} \, years$. This leads to a present upper bound of

$$[\#operations] \approx \rho \cdot c^5 \cdot t^4 \leq 10^{120} \qquad (15)$$

Lloyd further refines this to read

$$\#operations = \frac{4E}{\hbar} \cdot \left(t_1 - \sqrt{t_1 t_0}\right) \approx \left(t_{Final}/t_P\right) \leq 10^{120} \qquad (16)$$

We assume that $t_1$ = final time of physical evolution, whereas $t_0 = t_P \sim 10^{-43}$ seconds and that we can set an energy input by assuming, in early universe conditions, that $N^+ \neq \varepsilon^+ \ll 1$, and $0 < N^+ < 1$. So that we are looking at a graviton-burst-supplied energy value of

$$E = (V_{4-Dim}) \cdot \left[ \rho_{Vac} = \frac{\Lambda}{8\pi G} \right] \sim N^+ \cdot \left[ \rho_{graviton} \cdot V_{4-vol} \approx \hbar \cdot \omega_{graviton} \right] \qquad (17)$$

Furthermore, assuming the initial temperature is within the range of $T \approx 10^{32} - 10^{29}$ Kelvin, we have a Hubble parameter defined along the route specified by Lloyd. This is in lieu of time $t = 1/H$, a horizon distance defined as $\approx c/H$, and a total energy value within the horizon as

Energy (within the horizon) $\approx \rho_C \cdot c^3 / (H^4 \cdot \hbar) \approx 1/(t_P^2 \cdot H)$ \qquad (18)

And this for a horizon parameter Lloyd (2002) defines as:

$$H = \sqrt{8\pi G \cdot [\rho_{crit}]/3 \cdot c^2} \qquad (19)$$

And a early universe

$$\rho_{crit} \sim \rho_{graviton} \sim \hbar \cdot \omega_{graviton} / V_{4-Vol} \qquad (20)$$

Then

$$\# operations \approx 1/[t_P^2 \cdot H] \approx \sqrt{V_{4-Vol}} \cdot t_P^{-2} / \sqrt{8\pi G \hbar \omega_{graviton}/3c^2}$$
$$\approx [3\ln 2/4]^{4/3} \cdot [S_{Entrophy}/k_B \ln 2]^{4/3} \qquad (21)$$

### Claim 3: The Number of Allowed Operations In the Evolution of the Universe Specifies a Relationship Between an Evaluated Volume for Space-time and Upper Limits of Released Relic-Graviton Frequencies.

This is proved by appealing to Eqn. (21) above. Next, we will examine the existence of certain symmetries in the scalar field itself.

### FORMATION OF THE SCALAR FIELD, BIFURCATION RESULTS

Start with Padamans's formulas:

$$V(t) \equiv V(\phi) \sim \frac{3H^2}{8\pi G} \cdot \left(1 + \frac{\dot{H}}{3H^2}\right) \qquad (22)$$

$$\phi(t) \sim \int dt \cdot \sqrt{\frac{-\dot{H}}{4\pi G}} \qquad (23)$$

If $H = \dot{a}/a$, Eqn. (50) gives us zero scalar field values at the beginning of quantum nucleation of a universe. At the point of accelerated expansion (due to the final value of the cosmological constant), it also gives an accelerating value of the cosmological scale-factor expansion rate. We justify this statement by using early-universe expansion models, which have $a(t_{INITIAL}) \sim e^{H \cdot t}$. This leads to the derivative of $H = \dot{a}/a$ going to zero. This is similar to present-time development of the scalar factor along the lines of $a(t_{later}) \sim e^{(\Lambda[present-day]t)}$, also leading to the derivative of $H = \dot{a}/a$ going to zero. When both situations occur, we have the scale factor $\phi = 0$. Between initial and later times, the scale factor no longer has exponential time dependence, due to it growing far more slowly, leading to $\phi \neq 0$.

Both regimes as specified by Eqn. (50) above lead to zero values for a quintessence scalar field. But it does not stop there. We will show later that in actuality, the scalar field likely damps out far before the CMBR barrier value of expansion when Z = 1100, about 380,000 to 400,000 years after the big bang.

## Claim 4: We Observe That The Scalar Field $\phi(t)$ Is Zero At The Onset Of the Big Bang, And Also Is Zero During the Present Cosmological Era.

This scalar "quintessence" field is non zero in a brief period of time right after the inflationary era."

We show this by noting that in Eqn (22), the time derivative of $H = \dot{a}/a$ goes to zero when both the scale factors $a(t_{INITIAL}) \sim e^{H \cdot t(initial)}$, and $a(t_{later}) \sim e^{\Lambda[present-day]t(later)}$. The exponential scale factors in both cases (the initial inflationary environment and the present era) lead to the time derivative of the $H = \dot{a}/a$ expression in Eqn. (23) going to zero.

**Sub point to claim 4:** The existence of two zero values of the scalar field $\phi(t)$ at both the onset and at a later time implies a bifurcation behavior for modeling quintessence scalar fields. This is due to the non-zero $\phi(t)$ values right after the initiation of inflation.

How do we construct high-frequency gravity waves from all of this? Note that in Eqn. (21) above, that we have the existence of a denominator of the right hand side of the equation with the square of the Planck's time. It so happens that inflation is characterized with a rapid buildup of space-time volume. Dr. Smoot (2007) at the "D.Chalonge" colloquia in Paris specified initial computational bits of information transferred on the order of $10^8$ to $10^{10}$ bits of computation, expanding up to $10^{120}$ shortly after the initiation of the big bang itself. This creation of additional degress of freedom is in tandem with breaching the scale factor discontinuity mentioned in Claim 2 above. When we get to this regime of scale factor discontinuity, we get into the physics discussed by the claim given below.

## Claim 5: Unless the frequency $\omega_{graviton}$ in Eqn. (21) becomes large (~ $10^3$ Hz. or higher), the number of operations could effectively go to $10^{1000}$ or higher.

How do we show this? One would need to have very large gravitational frequency range, with high frequency gravity waves, in order to break the effects of a tiny Planck time interval $t_P^{-2} \sim 10^{86} \sec^{-2}$ put in the number of operations. So that instead of Eqn. (21) bounded by $10^{120}$, as the volume increased, the number of degrees of freedom of operations could become almost infinite.

This last claim--combined with the discussion right after Eqn. (11) above re the initial "drum head" model for a bounded region of space bracketed by causal discontinuity regions--constitutes our working model of an information-based model of cosmology that we expect will yield falsifiable experimental criteria.

## SMOOT'S INFORMATION THEORY/COSMOLOGY CONCLUSIONS AT PARIS COLLOQUIUM (2007 "D.CHALONGE" SCHOOL )

At the "D.Chalonge" school presentation done by Dr. Smoot (2007), he stated the following information-theory processing bits levels, which are due to different outlined physical processes. The following is Dr. Smoot's preliminary analysis of information content in the observable universe:

1) Physically observable bits of information possibly generated in the Universe: $10^{180}$
2) Holographic principle-allowed bits (states) in the evolution (development) of the Universe: $10^{120}$
3) Initially available bits (states) given to us to work with at the onset of the inflationary era: $10^{10}$
4) Observable bits of information present due to quantum/statistical fluctuations: $10^{8}$

Our guess is as follows: the thermal flux implied by the existence of a wormhole accounts for perhaps $10^{10}$ bits of information. These could be transferred via a wormhole solution from a prior universe to our present, per Eqn. (4) above. So there could be perhaps $10^{120}$ minus $10^{10}$ bits of information temporarily suppressed during the initial bozonification phase of matter right at the onset of the big bang itself. Then the degrees of freedom of our initial cosmological environment dramatically dropped during the beginning of the descent of temperature from about $T \approx 10^{32} \, Kelvin$ to at least three orders of magnitude. This drop in temperature occurs as we move out from an initial red shift of $z \approx 10^{25}$ to $z \approx 10^{25}$ to a far smaller value of the red shift,

$T \approx \sqrt{\varepsilon_V} \times 10^{28} \, Kelvin \sim T_{Hawkings} \cong \frac{\hbar \cdot H_{initial}}{2\pi \cdot k_B}$, as outlined by N. Sanchez (2007) at the "D.Chalonge" colloquium. A good guess as to what is going on is embodied in figure 28.1 of the book by Volovik (2003), offering a condensed matter analogy to current cosmology, where the formation of topological cosmic defects according to the Kibble –Zurek hypothesis may be in tandem with the growth of cosmological bits of information from a low number of $10^{10}$ to $10^{120}$ today. This is similar to the growth in baryons of up to $10^{80}$ in the modern era. The formation of complexity of structure as given in figure 28.1, cited above, occurs at the same time as the bit complexity reaches its present value of nearly $10^{120}$. This structure growth can be seen in vortex formation of a micro "big bang," with a low value of $10^{10}$ bits of information for the "complexity" of structure of the universe at or above a critical temperature $T \sim 10^{32} \, Kelvin$ for cosmological material just exiting the wormhole, to nearly $10^{120}$ bits of information for the "complexity" of the universe, which would be for temperatures $T \sim 10^{28} \, Kelvin$. Volovik's figure presents this in terms of a symmetry-breaking phase transition, which is similar to current cosmological models now in active development.

## CONCLUSION

So far, what we have established is a working model for an information theory based model of cosmological evolution with a lot of symmetry arguments thrown in. The approach is novel, leading to a new way of looking at CMBR space/volume and what it is, relative to bits of "information" computed during the course of cosmological evolution.

As for future research, we should delineate in more detail what would be transferred--possibly by entanglement information transfer from a prior universe to our own. We should also understand how additional bits of information came to be in the present Universe. All of this would tie in with an accurate physical understanding of the points raised in section V above. We would like to see--partly using the rich lore on liquid helium as outlined by Kopik (1993)--if there is a way to experimentally determine if the growth and the relative increase in structure and bits of "information" are in some sense connected with a cosmological equivalent to the vortex reconnection process outlined in liquid helium experiments. Our guess is that there is actually some symmetry-breaking transition equivalent in early universe cosmology that conceivably could be experimentally duplicated.

In particular, the author is convinced that the fifth claim as given above is fundamental physics, and that as we have a growing volume during inflation, this needs to be investigated--hopefully in ways that lead to falsifiable computational quantum computer models of the universe, and to their connections to initial inflation physics.

## ACKNOWLEDGMENTS

The author would like to acknowledge the contributions of Paul Murad in reviewing the physics and Amara D. Angelica for stimulating discussions of the concepts and copyediting.